\documentclass[5p,times]{elsarticle}

\biboptions{numbers,sort&compress}

\usepackage{mathtools}
\usepackage{mathrsfs}
\usepackage{accents}
\usepackage{algorithm}
\usepackage{algorithmicx}
\usepackage{algcompatible}
\usepackage{algpseudocode}
\usepackage{color}
\usepackage{comment}

\definecolor{purple}{rgb}{0.6, 0.4, 0.8}
\usepackage{cleveref}

\journal{Photoacoustics}

\newcommand{\FullEqRef}[1]{\textcolor{black}{Eq.~\eqref{#1}}}
\newcommand{\FullFigRef}[1]{\textcolor{black}{Fig.~\ref{#1}}}
\newcommand{\off}[1]{}

\begin{document}

\begin{frontmatter}

\title{Optoacoustic Model-Based Inversion Using Anisotropic Adaptive Total-Variation Regularization}

\author{Shai Biton$^1$, Nadav Arbel$^1$ ,Gilad Drozdov, Guy Gilboa, Amir Rosenthal*}
\address{Andrew and Erna Viterbi Faculty of Electrical Engineering, Technion
– Israel Institute of Technology, Technion City 32000, Haifa, Israel}
\fntext[myfootnote]{These authors contributed equally to this work}


\author[mysecondaryaddress]{A. Rosenthal\corref{mycorrespondingauthor}}
\cortext[mycorrespondingauthor]{Correspondence author}
\ead{amir.r@technion.ac.il}

\begin{abstract}
In optoacoustic tomography, image reconstruction is often performed with incomplete or noisy data,
leading to reconstruction errors. Significant improvement in reconstruction accuracy may be achieved
in such cases by using nonlinear regularization schemes, such as total-variation minimization and
$L_1$-based sparsity-preserving schemes. In this paper, we introduce a new framework for optoacoustic
image reconstruction based on adaptive anisotropic total-variation regularization, which is more
capable of preserving complex boundaries than conventional total-variation regularization. The new scheme is demonstrated in numerical simulations on blood-vessel images \textcolor{black} {as well as on experimental data} and is shown to be more capable than the total-variation-$L_1$ scheme in enhancing image contrast. 
\end{abstract}

\begin{keyword}
optoacoustic imaging, total variation, inversion algorithms, model-based reconstruction
\end{keyword}

\end{frontmatter}

\section{Introduction}
Optoacoustic tomography (OAT) is a hybrid imaging modality capable of visualizing optically absorbing structures with ultrasound resolution at tissue depths in which light is fully diffused \cite{ntziachristos2010going:33,wang2012photoacoustic:4, ntziachristos2010molecular:1,taruttis2015advances:2,wang2009multiscale:3}. The excitation in OAT is most often performed via high-energy short pulses whose absorption in the tissue leads to the generation of acoustic sources via the process of thermal expansion. The acoustic signals from the sources are measured over a surface that partially or fully surrounds the imaged object and used to form an image of the acoustic sources, which generally represents the local energy absorption in the tissue \cite{rosenthal2013acoustic:5}. Since hemoglobin is one of the strongest absorbing tissue constituents, optoacoustic images often depict blood vessels and blood-rich organs, such as the kidneys \cite{buehler2010video:34} and heart \cite{dean2015high:35}.

Numerous algorithms exist for reconstructing optoacoustic images from measured tomographic data \cite{rosenthal2013acoustic:5}. \textcolor{black}{In several imaging geometries, analytical formulae exist that may be applied directly on the measured data in either time \cite{xu2004time,xu2005universal:6} or frequency domain \cite{schulze2011use:7} to yield an exact reconstruction. The popularity of analytical formulae may be attributed to the simplicity of their implementation and low computational burden \cite{Rosenthal:18}} However, analytical algorithms  are not exact for arbitrary detection surfaces or detector geometries and lack the possibility of regularizing the inversion in the case of noisy or incomplete data. In such cases, it is often preferable to use model-based algorithms, in which the relation between the image and measured data is represented by a matrix whose inversion is required to reconstruct the image.

In the last decade, numerous regularization approaches have been demonstrated for model-based image reconstruction. The most basic approach is based on energy minimization and includes techniques such as Tikhonov regularization \cite{calvetti2000tikhonov:14} and  truncated singular-value decomposition \cite{yuan2005imaging:36}. In these techniques, a cost function on the image or components thereof is used to avoid divergence of the solution in the case of missing data, generally without making any assumptions on the nature of the solution. More advanced approaches to regularization exploit the specific properties of the reconstructed image. Since natural images may be sparsely represented when transformed to an alternative basis, e.g. the wavelet basis, using nonlinear cost functions that promote sparsity in such bases may be used for denoising and image reconstruction from missing data \cite{han2015sparsity:24,provost2009application:18,guo2010compressed:19,liang2009compressed:20,sun2011photoacoustic:21,meng2012compressed:22,liu2012compressed:23}. In images of blood vessels, in which the boundaries of the imaged structures may be of higher importance than the texture in the image, total-variation (TV) minimization has been shown to enhance image contrast and reduce artifacts \cite{wang2011limited:15,yao2011enhancing:16,yao2011photoacoustic:17}

In TV regularization, the cost function \textcolor{black}{regularizer is} the $L_1$ norm of the image gradient, which is generally lower for images with sharp, yet very localized, variations than for images in which small variations occur across the entire image. Therefore TV regularization enhances boundaries and reduces texture, where over-regularization may lead to almost piecewise constant images, which are often referred to as cartoon-like. While TV regularization is capable of accentuating the boundaries of imaged objects, it does not treat all boundaries the same. In particular, boundaries with short lengths will lead to a lower TV cost function than boundaries with long lengths. Thus, complex, non-convex boundaries may be rounded by TV regularization to the closest convex form.
\textcolor{black}{Recently, Wang  \textit{et al.} have shown that if the directionality of the TV functional is adapted to the image features, TV regularization may be applied for optoacoustic reconstruction of objects with non-convex boundaries without distorting the boundaries \cite{Wang}.} 


\looseness=-1
In this paper, we demonstrate a new regularization framework for model-based optoacoustic image
reconstruction that overcomes the limitations of TV regularization and is compatible with objects
with complex non-convex boundaries. In our \textcolor{black}{scheme}, an adaptive anisotropic total variation (A$^2$TV) cost function is used, in which the cost function is determined by the specific geometry of the imaged objects \cite{aatv}. In particular, the A$^2$TV cost function wishes to minimize the total variation in directions that are orthogonal to the boundary of the object. \textcolor{black}{In contrast to \cite{Wang}, where the boundaries were calculated using geometrical considerations limited to 2D images, the A$^2$TV framework developed in this work is based on eigenvalue decomposition of the image structure tensor, which may be applied in higher dimensions. 
The proposed formalism in the current study is } \textcolor{black}{  
based on a recent work by part of the authors which is concerned with nonlinear spectral analysis of the A$^2$TV functional \cite{aatv}. The work of Ref. \cite{aatv} examines  shapes which are perfectly preserved under A$^2$TV regularization. Earlier works concerning TV have shown that only convex rounded shapes of low curvature are preserved \cite{tvFlowAndrea2001}. For A$^2$TV, however, it is shown in \cite{aatv} that a parameter controlling the local extent of directionality is directly related to the degree of convexity (in the sense of \cite{peura}) and to the curvature magnitude of structures which are preserved. Thus, with appropriate parameters, long vessels of complex-nonconvex structure can be better regularized, keeping the original structure intact.
}

The performance of A$^2$TV regularization was tested numerically in this work in numerical simulations for complex images of blood vessels and \textcolor{black}{experimentally on 2D phantoms}. The simulations were performed for the cases of noisy data and missing data and compared to unregularized reconstructions as well as to TV-$L_1$ reconstructions \cite{han2015sparsity:24}. \textcolor{black}{ In both the numerical and experimental reconstructions, A$^2$TV significantly increased the image contrast and was more capable than TV-$L_1$ in preserving non-convex structures when strong regularization was performed. In the experimental reconstructions, A$^2$TV achieved a higher level of contrast enhancement of weak structures than the one achieved by TV-$L_1$. }



The rest of the paper is organized as follows: in Section 2 we give the theoretical background for OAT image reconstruction. Section 3 introduces the framework of A$^2$TV and the A$^2$TV algorithm for OAT image reconstruction developed in this work. \textcolor{black}{The simulation results are given in Section 4, while the experimental ones are given in Section 5. We conclude the paper with a Discussion in Section 6.} 

\section{Optoacoustic image reconstruction}
\subsection{The forward problem}
The acoustic waves in OAT are commonly described by a pressure field $p(\mathbf{r},t)$ that fulfills the following wave equation\cite{kruger1995photoacoustic:26}:
\begin{equation}
\frac{\partial^2p(\mathbf{r}, t)}{\partial t^2}-c^2 \nabla^2 p(\mathbf{r}, t) = \Gamma H_r (\mathbf{r})\frac{\partial H_t (t)}{\partial t}
\end{equation}
where $c$   is the speed of sound in the medium, $t$ is time, $\mathbf{r} =(x,y,z)$  denotes position in 3D space, $p(\mathbf{r},t)$ is the generated pressure, $\Gamma$ is the Gr{\"u}neisen parameter, and $H_r (\mathbf{r}) H_t (t)$  is the energy per unit volume and unit time. The spatial distribution function of energy deposited in the imaged object, $H_r (\mathbf{r})$,  is referred to in the rest of the paper as the optoacoustic image. 

The analysis of (1) for an optoacoustic point source at $\mathbf{r}'$, i.e. $H_r (\mathbf{r})=\delta(\mathbf{r-r'})$ , may be performed in either  time or frequency domain. In the time domain, a short-pulse excitation $H_t (t)= \delta(t)$ is used. In this case, the solution to \textcolor{black}{Eq.} (1) is given by \cite{blackstock2000fundamentals:27}
\begin{equation}
p_\delta (\mathbf{r},t) = \frac{\Gamma}{4\pi c}\frac{\partial}{\partial t}\frac{\delta(|\mathbf{r-r'}|-ct)}{|\mathbf{r-r'}|}.
\label{eq:delta}
\end{equation}
\textcolor{black}{For a general image $H_r (\mathbf{r})$, the solution for $p (\mathbf{r},t)$ may be obtained by convolving $H_r (\mathbf{r})$ with the expression in \FullEqRef{eq:delta}, which yields:
\begin{equation}
p (\mathbf{r},t) = \frac{\Gamma}{4\pi c}\frac{\partial}{\partial t}\int_{|\mathbf{r-r'}|=ct}\frac{H_r (\mathbf{r'})}{|\mathbf{r-r'}|}.
\label{eq:p_int}
\end{equation}}

\begin{figure}[t]
\centering
    \includegraphics[height=0.2\textwidth]{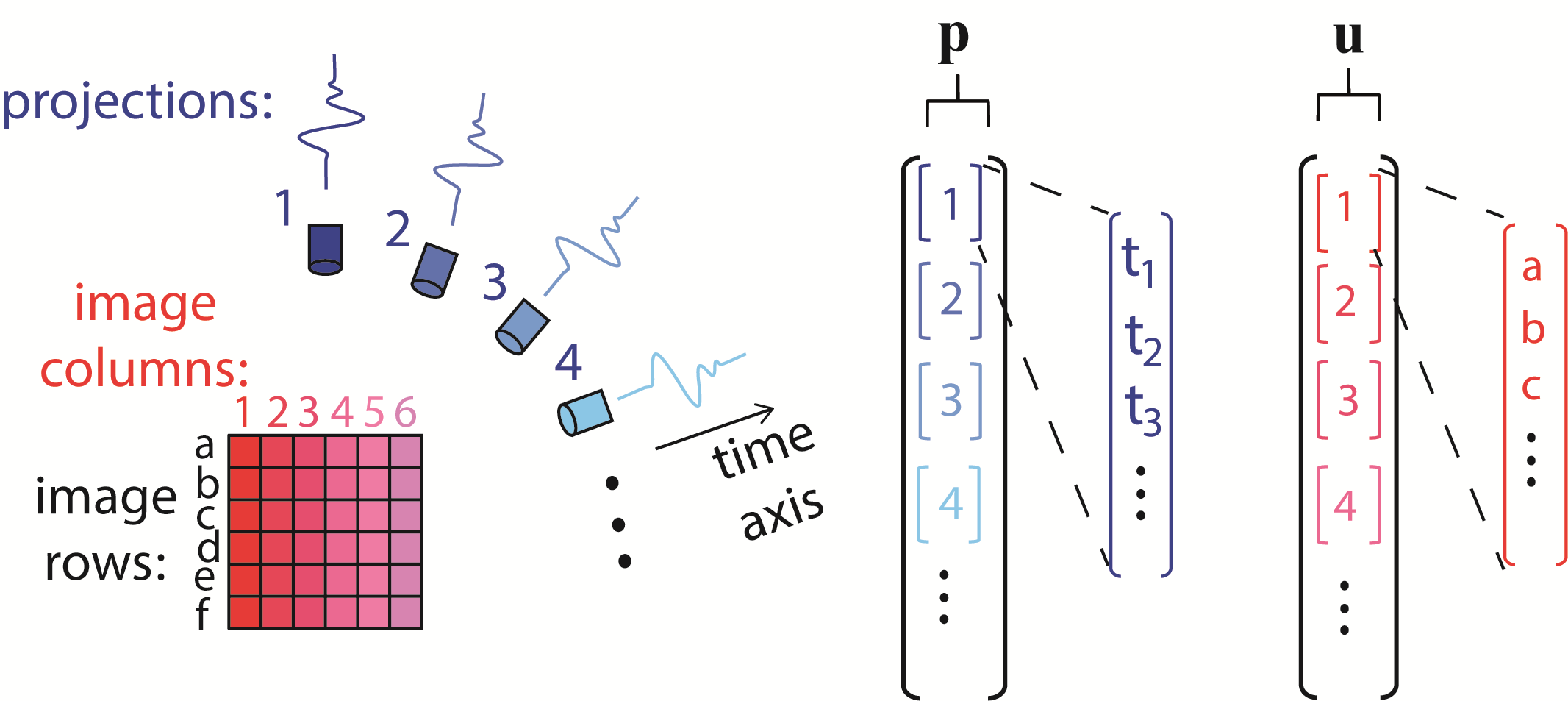}
  \caption{\textcolor{black}{An illustration of the structure of vectors $\mathbf{p}$ and $\mathbf{u}$ used in the matrix construction in \FullEqRef{eq:p=Mu}.}}
  \label{fig:matrix_construction}
\end{figure}

Since the relation between the measured pressure signals, or projections, and the image is linear, it may be represented by a matrix relation in its discrete form:
\begin{equation}
\mathbf{p}=\mathbf{Mu}
\label{eq:p=Mu}
\end{equation}
where $\mathbf{p}$ and  $\mathbf{u}$  are vector representations of the acoustic signals and originating image respectively, and $\mathbf{M}$ is the model matrix that represents the operations in \FullEqRef{eq:p_int}. \textcolor{black}{In our work, the images is given on a two dimensional grid, and the measured pressure signals are also two dimensional, where one dimension represents the projection number, and the other time. An illustration of the image grid and projection and their respective mapping to the vectors $\mathbf{u}$ and $\mathbf{p}$, is shown in \FullFigRef{fig:matrix_construction}. Briefly, the vector $\mathbf{u}$ is divided into sub-vectors, each representing the image values for a given column, whereas the vector $\mathbf{p}$ is divided to sub-vectors, each of which represents the time-domain pressure signal for a given location of the acoustic detector.} 

\textcolor{black}{The \textit{i}th column of the matrix $\mathbf{M}$ represents the set of acoustic signals generated by a pixel corresponding to the location of the \textit{i}th entry in the vector $\mathbf{u}$. Accordingly, to calculate the matrix $\mathbf{M}$ one needs to define time domain signal for a given detector location expected for a discrete pixel. Since the operations in \FullEqRef{eq:p_int} relate to continuous, rather than discrete images, one first needs to define the continuous representation of a single pixel, and then calculate its respective time-domain signals. For example, in \cite{paltauf2002iterative} it was assumed that the image value was constant over each of the square pixels, leading to an image $H_r (\mathbf{r})$ that is piece-wise constant. While simple to implement, a simple piece-wise uniform model for $H_r (\mathbf{r})$ includes discontinuities that lead to significant numerical errors owing to the derivative operation in \FullEqRef{eq:p_int}. In the current work, we use the model of  \cite{rosenthal2010fast:8}, in which the image $H_r (\mathbf{r})$ represented by a linear interpolation between its grid points.}     

\subsection{The inverse problem}
While several approaches to OAT image reconstruction exist, we focus herein on image reconstruction
within the discrete model-based framework described in the previous sub-section, which involves
inverting the matrix relation in~\FullEqRef{eq:p=Mu} to recover $\mathbf{u}$ from $\mathbf{p}$. The most
basic method to invert~\FullEqRef{eq:p=Mu} is based on solving the following optimization problem:  
\begin{equation}
\mathbf{u}^*=\arg\min||\mathbf{p}-\mathbf{Mu}||_2^2,
\label{eq:argmin pmu}
\end{equation}
where $\mathbf{u}^*$ is the solution and $||\cdot||_2$ is the $L_2$ norm. A unique solution
to~\FullEqRef{eq:argmin pmu} exists, which is given by the Moore--Penrose inverse:
\begin{equation}
\mathbf{u}^*=(\mathbf{M^TM})^{-1}\mathbf{M}^T \mathbf{p}.
\end{equation}
Alternatively,~\FullEqRef{eq:argmin pmu} may be solved via iterative optimization algorithms. In particular, since the matrix $\mathbf{M}$ is sparse, efficient inversion may be achieved by the LSQR algorithm \cite{paige1982lsqr:28}

In many cases, the measured projection data $\mathbf{p}$ is insufficient to achieve a high-quality reconstruction of $\mathbf{u}$ that accurately depicts its morphology. For example, when the density or coverage of the projections is too small, the matrix $\mathbf{M}$ may become ill-conditioned, leading to significant, possibly divergent, image artifact. In other cases, $\mathbf{M}$ may be well conditioned, but the measurement data may be too noisy to accurately recover $\mathbf{u}$. In both these cases, regularization may be used to improve image quality by incorporating previous knowledge on the properties image in the inversion process.     

One of the simplest forms of regularization is Tikhonov regularization, in which an additional cost function is added \cite{calvetti2000tikhonov:14}: 
\begin{equation}
    \mathbf{u}^*=\arg\min{\|\mathbf{p}-\mathbf{Mu}\|_2^2+\lambda\|\mathbf{Lu}\|_2^2}
\end{equation}
where $\mathbf{L}$ is weighting matrix. In the simplest form of Tikhonov regularization, $\mathbf{L}$ is equal to the identity matrix, i.e. $\mathbf{L}=\mathbf{I}$, thus putting a penalty on the energy of the image. The value of the regularization parameter $\lambda>0$ controls the tradeoff between fidelity and smoothness, where over-regularization may lead to the smearing of edges and texture in the image.

An alternative to the energy-minimizing cost function of Tikhonov are sparsity-maximizing cost functions. Since natural images may be sparsely represented in an alternative basis, e.g. the wavelet basis, a cost function that promotes sparsity may reduce reconstruction errors. Denoting the transformation matrix by $\mathbf{\Phi}$, one wishes that $\mathbf{\Phi u}$ be sparse, i.e. that most of its entries be approximately zero. In practice, sparsity is often enforced by using the $L_1$ norm because of its compatibility of optimization algorithms \cite{provost2009application:18,guo2010compressed:19,liang2009compressed:20,sun2011photoacoustic:21,meng2012compressed:22,liu2012compressed:23}. Accordingly, the inversion is performed by solving the following optimization problem:      
\begin{equation}
    \mathbf{u}^*=\arg\min{||\mathbf{p}-\mathbf{Mu}||_2^2+\mu\|\mathbf{\Phi}\mathbf{u}\|_1},
\end{equation}
where $\mu>0$ is the regularization parameter, controlling the tradeoff between sparsity and signal fidelity. When over regularization is performed, compression artifacts may appear in the reconstructed image. 

Sparsity may be enforced not only on alternative representations of the image, but also on image variations. The discrete TV cost function approximates the $l_1$ norm on the image gradient, and is given by 
\begin{equation}
    \|\mathbf{u}\|_{TV}=\sum_{n}{\sqrt{|u_{x,y}^n-u_{x-1,y}^n|^2+|u_{x,y}^n-u_{x,y-1}^n|^2}},
    \label{eq:utv}
\end{equation}
where $u_{x,y}^n$ is the $n$th entry of the vector $\mathbf{u}$, and the subtraction of "1" to $x$ or $y$ in the subscript corresponds to an entry of $\mathbf{u}$ that is respectively shifted in relation to $u_{x,y}^n$ by one pixel in the $x$ or $y$ direction of the 2D image. The inversion using TV is thus given by \cite{wang2011limited:15,yao2011enhancing:16,yao2011photoacoustic:17}
\begin{equation}
\label{eq:inv_tv}
\mathbf{u}^*=\arg\min{||\mathbf{p}-\mathbf{Mu}||_2^2+\alpha\|\mathbf{u}\|_{TV}}
\end{equation}
where $\alpha>0$ is the regularization parameter. In the case of TV minimization, over-regularization may lead to cartoon-like images and rounding of complex boundaries into convex shapes.  

In some cases, it is beneficial to promote both sparsity of the image in an alternative basis and TV minimization. In such cases, the optimization problem is as follows \cite{han2015sparsity:24}: 
\begin{equation}
    \mathbf{u}^*=\arg\min{||\mathbf{p}-\mathbf{Mu}||_2^2+\mu\|\mathbf{\Phi}\mathbf{u}\|_1}+\alpha\|\mathbf{u}\|_{TV}.
    \label{eq:anotheruprob}
\end{equation}
\textcolor{black}{We will refer to the regularization described in \FullEqRef{eq:anotheruprob}, as TV-$L_1$ regularization.} 

\section{Adaptive Anisotropic Total Variation}


\subsection{The Functional}

\begin{figure}[t]
\centering
  \ifpdf
    \includegraphics[width=0.3\textwidth]{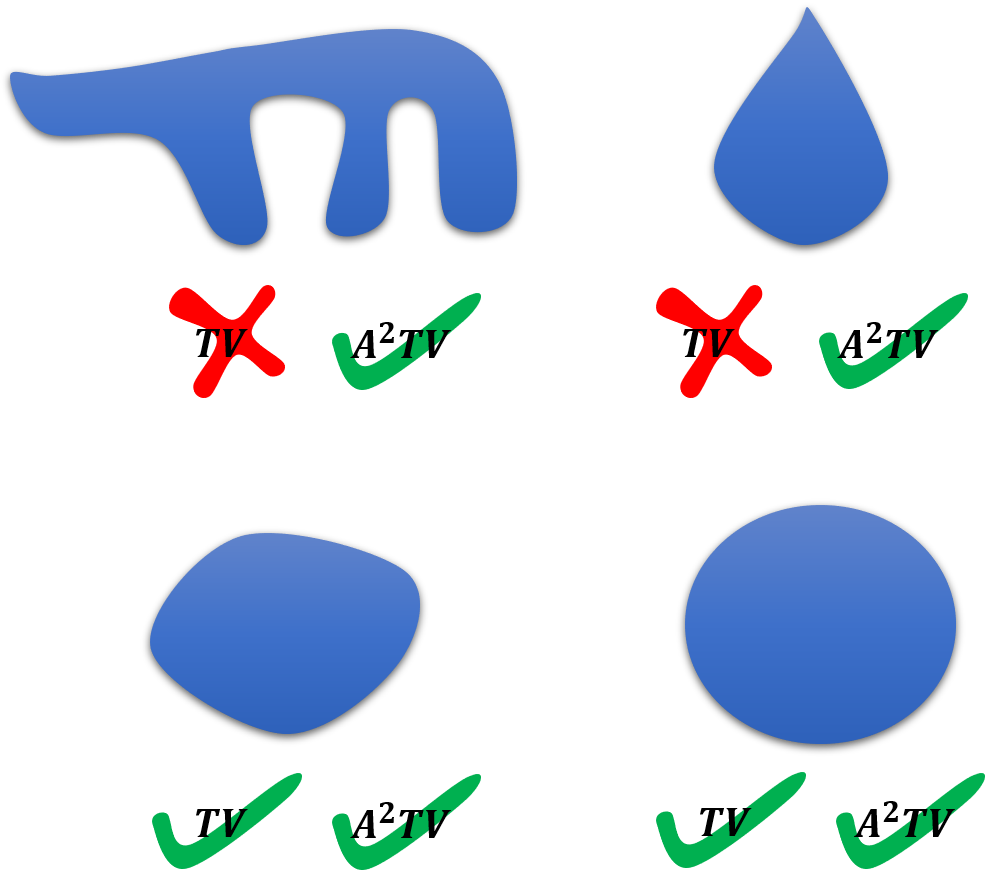}
  \else
    \includegraphics[width=0.3\textwidth]{Graphics/Eigen.png}
  \fi
  \caption{An illustration of sets which are stable for TV and A$^2$TV – notice A$^2$TV admits non-convex and highly curved functions, including ones which resemble arteries.} 
  \label{fig:eigen}
\end{figure}

We would like to use a regularizer that is adapted to the image in such a way that it regularizes more along edges (level-lines of the image) and less across edges (in the direction of the gradient). This idea has been introduced for nonlinear scale-space flows by Weickert \cite{weickert1998anisotropic:31} in the anisotropic diffusion formulation. However there is no known functional associated with anisotropic diffusion, and it is therefore not trivial to include an anisotropic-diffusion operation in our inverse-problem formulation. 
A more recent study of Grasmair \textit{et al.} \cite{grasmair2010anisotropic:38} uses a similar adaptive scheme within a TV-type formulation.
In the study of \cite{aatv}, a comprehensive theoretical and numerical analysis was performed for adaptive-anisotropic TV (A$^2$TV). It was shown that stable structures can be non-convex and in addition can have high curvature on the boundaries. \textcolor{black}{ Illustrations of the stable sets characterize the TV and A$^2$TV regularizers are shown in \FullFigRef{fig:eigen}.}
 The degree of anisotropy directly controls the degree of allowed nonconvexity and the upper bound on the curvature. We adopt the formulation of  \cite{aatv}, in which the mathematical underpinnings of A$^2$TV are described in detail.  
 
Let the A$^2$TV functional be defined by
\begin{equation}
    \mathcal{J}_{A^2TV}(\mathbf{u})=\int_\Omega \|A(\mathbf{x})\nabla
    \mathbf{u}(\mathbf{x})\|_2d\mathbf{x}=\int_\Omega \|\nabla_A\mathbf{u}(\mathbf{x})\|_2d\mathbf{x}
    \label{AATV_func}
\end{equation}
where $\mathbf{x} =(x,y)\in\Omega$  denotes position in 2D space, $A(\mathbf{x})\in \mathbb{R}^{2\times 2}$ is a tensor or a spatially adaptive matrix and $\nabla_A= A(\mathbf{x})\nabla$ is an ``adaptive gradient''. This functional is convex and can be optimized by standard convex solvers given the tensor $A(\mathbf{x})$. We now turn to the issue of how this tensor is constructed to allow a good regularization of vessel-like structures.


\subsubsection{Constructing the tensor $A(\mathbf{x})$}
\begin{figure}[t]
\centering
    \includegraphics[width=0.25\textwidth]{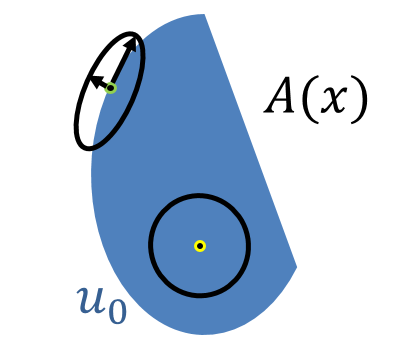}
  \caption{{ An illustration of the tensor $A(\mathbf{x})$.} At any point the tensor rotates and rescales the coordinate system in an image-driven manner. It assumes some approximation $u_0$ of the data exists. The tensor is designed such that lower regularization is applied across edges (top left ellipse) whereas in flat regions regularization is applied in an isotropic manner (bottom right circle). }
  \label{fig:Ax}
\end{figure}

We assume to have a rough approximation of the image to be reconstructed. This can be done, for
instance, by having an initial least-square non-regularized approximation or a standard TV
reconstruction, as in \FullEqRef{eq:inv_tv}. We thus have an initial estimation $\mathbf{u}_0$. \textcolor{black}{The tensor
$A(\mathbf{x})$ determines the principle and the secondary directions of the regularization at each point and their magnitude. The construction of $A(\mathbf{x})$ is performed using $\mathbf{u}_0$ according to the following principles: In regions in which $\mathbf{u}_0$ is relatively flat, i.e. $\nabla \mathbf{u}_0$ almost vanishes, $A(\mathbf{x})$ should resemble the identity matrix and have no preferred direction, thus leading to the conventional TV regularizer. In regions with dominant edges, $A(\mathbf{x})$ should capture the principle axes of the edge. See \FullFigRef{fig:Ax} for an illustration of $A(\mathbf{x})$.}

Mathematically, the tensor $A(\mathbf{x})$ is defined by adapting the eigenvalues of a smoothed structure tensor of a smoothed image $\mathbf{u}_{0;\sigma}$ (with a Gaussian kernel of standard deviation $\sigma$), defined by,
\begin{equation}
    \mathcal{J}_{\rho}(\nabla \mathbf{u}_{0;\sigma})=\kappa_\rho*\left(\nabla \mathbf{u}_{0;\sigma}\otimes\nabla \mathbf{u}_{0;\sigma}\right),
	\label{Eq:StructureTensor_generalcase}
\end{equation}
where $\kappa_\rho$ is a Gaussian kernel with a standard deviation of $\rho$, $*$ denotes an element-wise convolution and $\otimes$ denotes an outer product. \textcolor{black}{The Gaussian kernel's standard deviations $\sigma$ and $\rho$ are chosen in coordination with the noise level of the image and the smallest object resolution within the image. High standard deviation values might diminish the effect of small objects on the structure tensor while low values might lead to the unwanted scenario in which reconstruction errors have a significant effect on the structure tensor. The explicit expressions for $\mathcal{J}_{\rho}$ in the 2D and 3D cases are respectively given by 
\begin{equation}
\mathcal{J}_{\rho}(\mathbf{x}) = \begin{pmatrix} \kappa_\rho*u_{x;\sigma}^2(\mathbf{x}) & \kappa_\rho*u_{x;\sigma}(\mathbf{x})u_{y;\sigma}(\mathbf{x}) \\ \kappa_\rho*u_{x;\sigma}(\mathbf{x})u_{y;\sigma}(\mathbf{x}) & \kappa_\rho*u_{y;\sigma}^2(\mathbf{x}) \end{pmatrix}
    \label{eq:tensor2d}
\end{equation}
and
\begin{equation}
\mathcal{J}_{\rho}(\mathbf{x}) = \kappa_\rho* \begin{pmatrix} 
u_{x;\sigma}^2(\mathbf{x}) & u_{x;\sigma}(\mathbf{x})u_{y;\sigma}(\mathbf{x}) &
u_{x;\sigma}(\mathbf{x})u_{z;\sigma}(\mathbf{x}) &\\ 
u_{x;\sigma}(\mathbf{x})u_{y;\sigma}(\mathbf{x}) &
u_{y;\sigma}^2(\mathbf{x}) & 
u_{y;\sigma}(\mathbf{x})u_{z;\sigma}(\mathbf{x}) &\\ 
u_{x;\sigma}(\mathbf{x})u_{z;\sigma}(\mathbf{x}) &
u_{y;\sigma}(\mathbf{x})u_{z;\sigma}(\mathbf{x}) & u_{z;\sigma}^2(\mathbf{x}) \end{pmatrix}.
\end{equation}} 
 
The structure tensor matrix has eigenvectors corresponding to the direction of the gradient and tangent at each point $\mathbf{x}$; and eigenvalues corresponding to the magnitude of each direction. In order to preserve structure, we should change the relation between those eigenvalues so that for flat-like areas in the image we will smooth the image in an isotropic way, while for edge-like areas, we will perform more smoothing in the tangent direction rather the gradient one.
For 2D, we begin by looking in the eigen-decomposition of the structure-tensor,
\begin{equation}
\mathcal{J}_{\rho}=VDV^{-1}
\end{equation}
Where $V$ is a matrix whose columns, $v_1,v_2\in\mathbb{R}^2$, are the eigenvectors of $\mathcal{J}_{\rho}$, and $D$ is a diagonal matrix with eigenvalues in the diagonal,
\begin{equation}
D=\begin{pmatrix} \mu_1 & 0 \\ 0 & \mu_2 \end{pmatrix}; V=(v_1 | v_2)
\end{equation}
Assuming $\mu_1 \geq \mu_2$. \textcolor{black}{The spatially adaptive matrix $A(\mathbf{x})$, used in \FullEqRef{AATV_func}, is constructed from a modification of the eigenvalue matrix, denoted by $\tilde{D}$, as follows \cite{weickert1998anisotropic:37}:
\begin{equation}
A=V\tilde{D}V^{-1},
\end{equation}
where $\tilde{D}$ is given by 
\begin{equation}
\tilde{D}=\begin{pmatrix} c(\mu_1/\mu_{1,avg};k) & 0 \\ 0 & 1 \end{pmatrix}.
\label{eq:Dtilde}
\end{equation}
In \FullEqRef{eq:Dtilde}, $\mu_{1,avg}$ is the average value of $\mu_{1}$ across the image and $c(\cdot;\cdot)$ is a function  of two parameters defined as follows:}
\begin{equation}
c(s;k)= \left\{ \begin{array}{rcl}
1 & \mbox{,} & s\leq0 \\
1-exp(-\frac{c_m}{(\frac{s}{k})^m}) & \mbox{,} & s>0
\end{array}\right.
\end{equation}
\textcolor{black}{where the chosen values for the parameter are the ones recommended in Ref.  \cite{weickert1998anisotropic:31}: $c_m = 3.31488$ and $m = 4$. The parameter $k\leq 1$ determines which regions in the image will be regularized anisotropically and is chosen based on the desired level of anisotropy in the reconstructed image, as shown in Sections 4 and 5. In pixels in which $s \ll k$, i.e. $\mu_1/\mu_{1,avg} \ll k$, we will obtain $c \approx 1$, and $\tilde{D}$ will be reduced to the unitary matrix. Accordingly, for regions in which the image gradient is sufficiently small from the average image gradient, as regulated by $k$, the A$^2$TV functional (\FullEqRef{AATV_func}) is reduced to the standard, isotropic TV functional (\FullEqRef{eq:utv}). In the rest of the image, where $c$ is sufficiently smaller than 1, regularization is performed more strongly in the direction of the eigenvector $v_2$, i.e. the direction in which the image gradient is smaller, thus enhancing the anisotropy in those regions}

It is worth noting that in the 3D case, assuming $\mu_1 \geq \mu_2 \geq \mu_3$, the only modification to the analysis above is that $\tilde{D}$ accepts the following form:
\begin{equation}
\tilde{D}=\begin{pmatrix} c(\mu_1/\mu_{1,avg};k) & 0 & 0\\ 0 & c(\mu_2/\mu_{1,avg};k) & 0\\ 0 & 0 & 1 \end{pmatrix}.
\label{eq:D3D}
\end{equation}
\textcolor{black} {The structure of $\tilde{D}$ in \FullEqRef{eq:D3D} will be highly anisotropic for tube-like structures and will enforce variation-reducing regularization along the structure length (eigenvector $v_3$), while maintaining low regularization in the cross-section plane (eigenvectors $v_1$ and $v_2$).}


\off{
\subsection{Eigenfunctions of the TV and AATV functionals}
Eigenfunctions of functionals often describes the natural shapes and priors of which the functional forces as a regularizer of a given image. Each eigenfunction has its corresponding eigenvalue.
For the TV functional it has been shown that for a given eigenvalue $\lambda_{TV}$ the space of eigenfunctions is comprised of characteristic functions $\chi_C$ of convex sets $C$ defined as:  
\begin{equation}
\chi_C=\left\{ \begin{array}{rcl}
1 & \mbox{,} & \mathbf{x}\in C \\
0 & \mbox{,} & \mathbf{x}\notin C
\end{array}\right.
\end{equation}
Which obey the constraint of being convex and fulfilling the relationship:
\begin{equation}
\lambda_{TV}=\frac{Per(C)}{|C|}
\end{equation}
whereas AATV admits non-convex and highly curved shapes as an eigenfunctions with 
\begin{equation}
\lambda_{AATV}=a\frac{Per(C)}{|C|}
\end{equation}
Where $a=c(\mu_1;K)$ is constant since $u=\chi_C$.
Given a set C and its convex hull $D$, the set $C$ is an eigenfunction of the AATV functional if it fulfills
\begin{equation}
a\leq\frac{Per(C)}{Per(D)},\hspace{2mm}
\max_{\forall \mathbf{x}\in \partial C}\kappa(\mathbf{x})\leq\frac{1}{a^2}\frac{Per(C)}{|C|}.
\end{equation}
Which in the real world translates to choosing a small enough $a$ parameter to include as many eigenfunctions in the current AATV functional family, to adjust to most of the curved non convex boundaries of the objects in the image at hand,
}  

\subsection{Reconstruction based on A$^2$TV}
\textcolor{black}{The reconstruction based on the A$^2$TV minimizes the following functional:}
\begin{equation}
    \mathbf{u}^* = 
    \arg\min_{\mathbf{u}}\mathcal{J}_{A^2TV}(\mathbf{u})+\frac{\lambda}{2}\|\mathbf{M}\mathbf{u}-\mathbf{p}\|_2^2,
    \label{eq:lovelyuprob}
\end{equation} 
where $\mathbf{M}$ is the model matrix and $\mathbf{p}$ is the acoustic pressure wave. 
The solution of which is done by the modified Chambolle-Pock projection algorithm \cite{chambolle2011first:32} described in Appendix A.

In the process of minimization, the tensor $A$ is initialized as the identity matrix for all $\mathbf{x}\in\mathbb{R}^2$, which reduces the A$^2$TV energy to the TV one as it performs the diffusion isotropically. The tensor $A$ is then updated according to the initial solution $u_0$. This is repeated until numerical convergence is reached. 

We note that while the energy of the A$^2$TV is convex for a fixed tensor $A(\mathbf{x})$, it is not convex when $A(\mathbf{x})$ is adaptive and depends on the imaged object. Thereby,  we do not have a mathematical proof of convergence. Nonetheless, it has been shown both in our work \cite{aatv} and in Refs. \cite{weickert1998anisotropic:37,grasmair2010anisotropic:38} that heuristically, both the image $u$ and the tensor $A(\mathbf{x})$ converge.

\section{Numerical Simulations}

\begin{figure}[t]
\centering
  \includegraphics[height=0.33\textwidth]{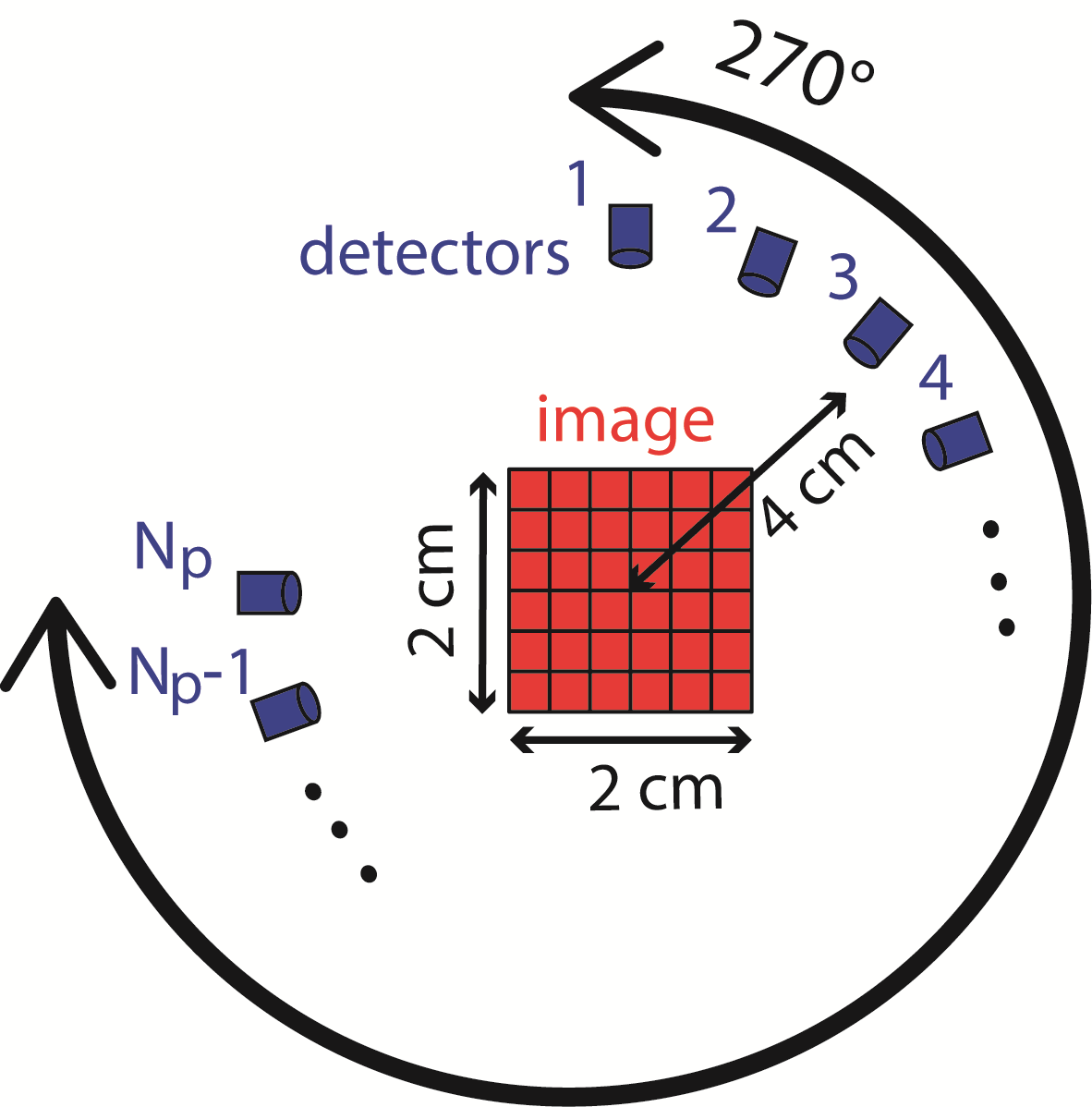}
  \caption{\textcolor{black}{An illustration of the image and detector geometry used in the numerical simulations and experimental measurements in Sections 4 and 5.}}
  \label{fig:simulation_illustration}
\end{figure}

\begin{figure}[t]
\centering
  \ifpdf
    \includegraphics[width=0.3\textwidth]{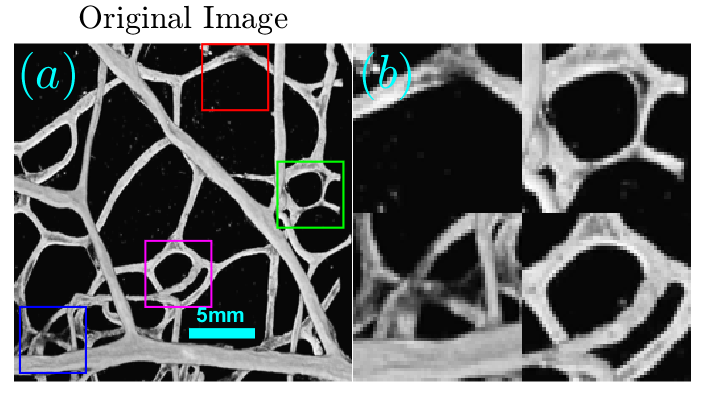}
  \else
    \includegraphics[width=0.3\textwidth]{Graphics/Optoacoustic_Artery_OriginalImage.png}
  \fi
  \caption{(a) The originating image on which all the reconstructions were performed and (b) a binary mask that was generated from it. The bottom panel (c-d) shows magnifications of 4 marked regions in the respective images in the top panel.}
  \label{fig:original_image}
\end{figure}

\textcolor{black}{In this section, we demonstrate the performance of A$^2$TV-based inversion for the circular detection geometry illustrated in \FullFigRef{fig:simulation_illustration}. The simulations were performed on a 2D vascular image of a mouse retina,  obtained via confocal microscopy. The vasculature image, shown in \FullFigRef{fig:original_image}a, was represented over a square grid with a size of 256$\times$256 pixels with pixel size of 0.1$\times$0.1 mm. The projections were simulated over a 270-degree
semi-circle with a radius of 4 cm that surrounded the object, in accordance with conventional OAT systems \cite{taruttis2015advances:2}. A magnification of four square regions of the image is shown in \FullFigRef{fig:original_image}b.} 


The reconstructions were performed using the conventional
$L_2$-based regularization-free approach~(\FullEqRef{eq:argmin pmu}) performed via LSQR, TV-$L_1$ regularization~(\FullEqRef{eq:anotheruprob}), and the proposed
A$^2$TV regularization~(\FullEqRef{eq:lovelyuprob}) method. 
\textcolor{black}{Since the scaling of the model matrix $\mathbf{M}$ (\FullEqRef{eq:p=Mu}) depends on the exact implementation of its construction \cite{rosenthal2010fast:8}, we normalized $\mathbf{M}$ by $\frac{1}{160}\sqrt{\|\mathbf{M}\|_\infty\|\mathbf{M}\|_1}$ to assure that the regularization parameters are independent of the scaling of $\mathbf{M}$.} \textcolor{black}{ To assess the quality of the reconstructions, we used the mean absolute distance (MAD) given by the following equation:
\begin{equation}
    \textnormal{MAD}=\frac{1}{N}\|\mathbf{u}_{\textrm{orig}}-\mathbf{u}^*\|_1,
    \label{eq:MAD}
\end{equation}}
\textcolor{black}{where $N$ is the number of pixels in the image.}


\textcolor{black}{Two cases were tested: In the first case, zero-mean Gaussian noise with a standard deviation of 0.6 times the maximum value of $\mathbf{p}$ was added to the projection. The number of projections was chosen to be 256, corresponding to the geometry found in state-of-the-art optoacoustic systems \cite{dima2014multispectral} and sufficient for the accurate reconstruction of the tested image in the noiseless case. In the second case, the number of projections was reduced to 32, which is half the number of projections used in low-end optoacoustic systems characterized by reduced lateral resolution \cite{dima2014multispectral}. Accordingly, 32 projections are insufficient for producing detailed optoacoustic images using conventional reconstruction techniques.} 
\textcolor{black}{In all the examples, the number of iterations was chosen to be sufficiently high to achieve convergence.}



\begin{figure}[t]
\centering
  \ifpdf
    \includegraphics[width=0.46\textwidth]{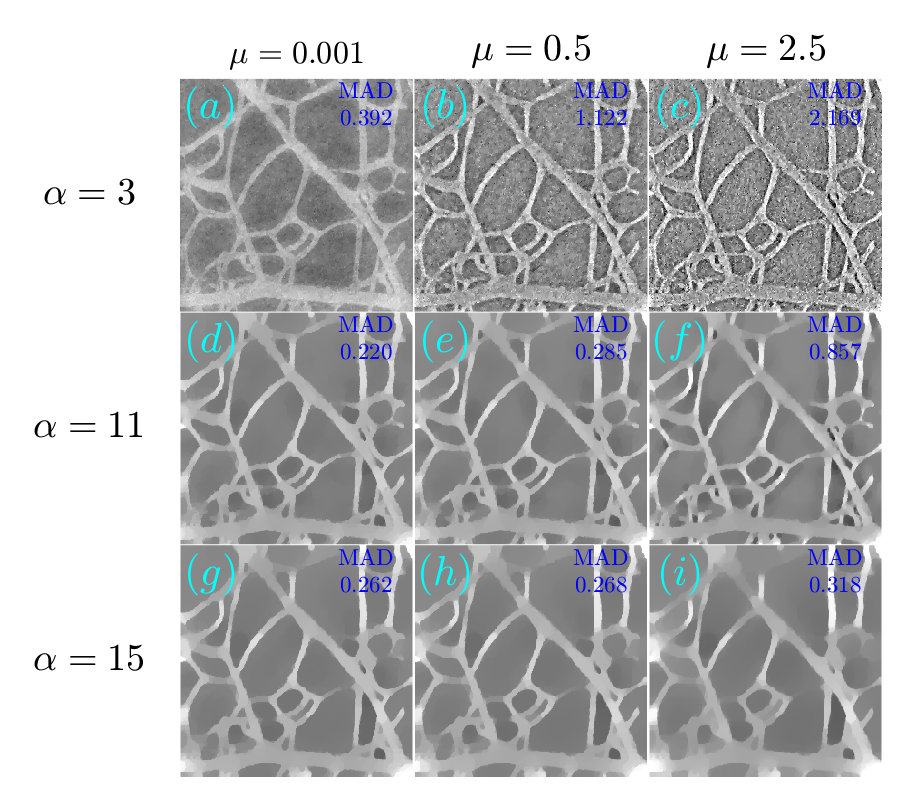}
  \else
    \includegraphics[width=0.46\textwidth]{Graphics/Optoacoustic_TVL1_256_noise.png}
  \fi
  \caption{\textcolor{black}{(a-i) The reconstruction of the image shown in \FullFigRef{fig:original_image}a for the case of additive Gaussian noise using different parameters for the TV-$L_1$ case. The reconstructions were performed with 3000 iterations.}}
  \label{fig:Recon_Params_Grid_TVL1_Gaussian-Noise}
\end{figure}

\begin{figure}[h]
\centering
  \ifpdf
    \includegraphics[width=0.46\textwidth]{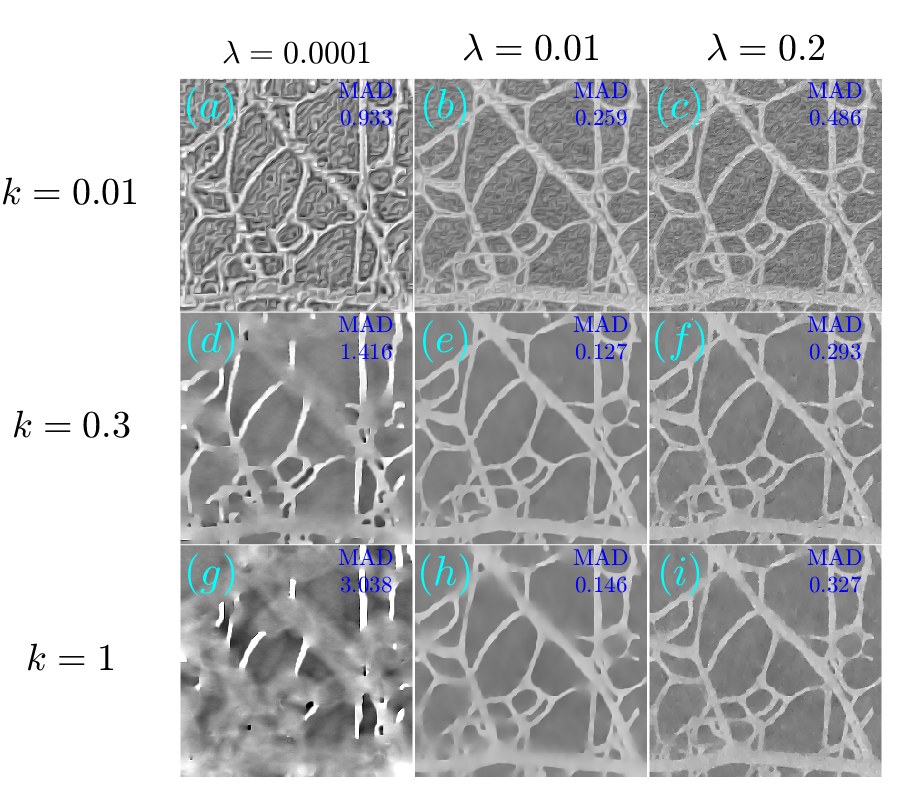}
  \else
    \includegraphics[width=0.46\textwidth]{Graphics/Optoacoustic_A2TV_256_noise.png}
  \fi
  \caption{\textcolor{black}{(a-i) The reconstruction of the image shown in \FullFigRef{fig:original_image}a for the case of additive Gaussian noise using different parameters for the A$^2$TV case. The reconstructions were performed with 3000 iterations.}}
  \label{fig:Recon_Params_Grid_A2TV_Gaussian-Noise}
\end{figure}

\begin{figure}[hp!]
\centering
  \ifpdf
    \includegraphics[width=0.46\textwidth]{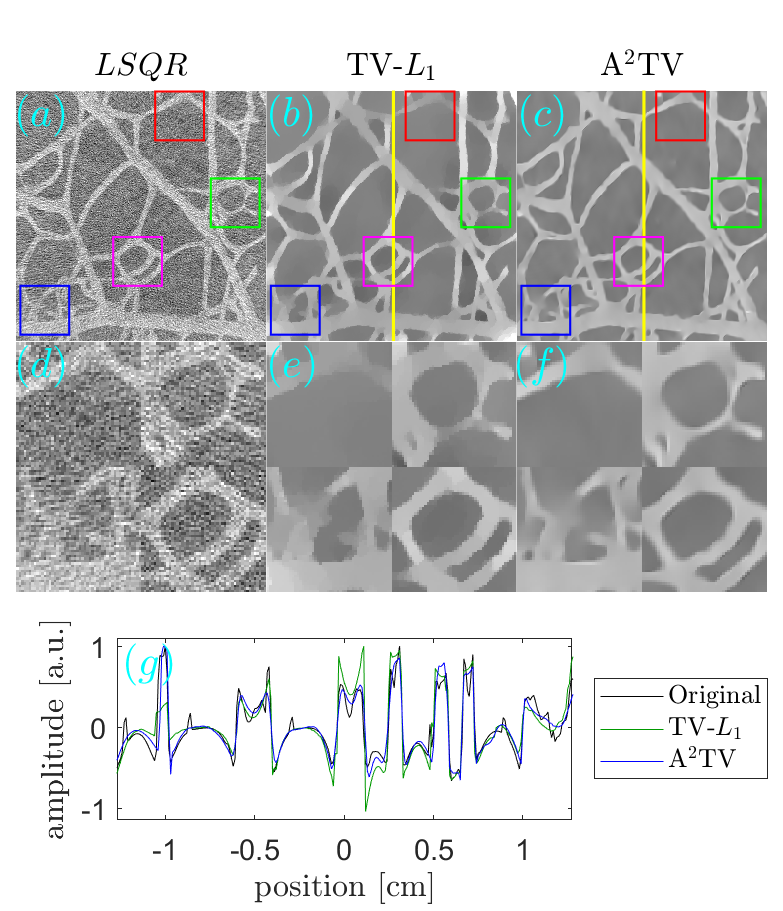}
  \else
    \includegraphics[width=0.46\textwidth]{Graphics/Optoacoustic_Hans_Comparison_noise_Noise_Example.png}
  \fi
  \caption{(a-c) The reconstruction of the image shown in \FullFigRef{fig:original_image}a for the case of additive Gaussian noise using (a) LSQR, (b) TV-$L_1$, and (c) A$^2$TV. The bottom panel (d-f) shows magnifications of 4 marked regions in the respective images in the top panel.  \textcolor{black}{Both the TV-$L_1$ and A$^2$TV reconstructions were produced using 3000 iterations.} \textcolor{black}{(g) A 1D slice of the TV-$L_1$ and A$^2$TV reconstructions, corresponding to the vertical lines on the top panel, in comparison to a 1D slice taken from the original image.}}
  \label{fig:gaussian noise}
\end{figure}

\textcolor{black}{\FullFigRef{fig:Recon_Params_Grid_TVL1_Gaussian-Noise} and \FullFigRef{fig:Recon_Params_Grid_A2TV_Gaussian-Noise} respectively show the reconstructions obtained using TV-$L_1$ and A$^2$TV, performed with 3000 iterations, for the cases of additive Gaussian noise. In both figures, 9 reconstructions are shown, corresponding to a scan in the regularization parameters. In the case of TV-$L_1$, $\mu$ and $\alpha$ represent the strength of the $L_1$ and TV regularization terms \FullEqRef{eq:anotheruprob}, whereas in the case of A$^2$TV, $\lambda$ represents the strength of the fidelity term in \FullEqRef{eq:lovelyuprob} with respect to the A$^2$TV term and $k$ determines the strength of the anisotropy, where lower values of $k$ correspond to higher anisotropy. \textcolor{black}{In the A$^2$TV reconstructions, the standard deviations of the smoothing kernels (Eq. \ref{eq:tensor2d}) were $\sigma=1.5$ pixels and $\rho=3$ pixels.} For all the reconstructions, the MAD values appear on the top-right corner of the image.  In \FullFigRef{fig:gaussian noise}, we show a comparison between the regularization-free LSQR reconstruction (\FullFigRef{fig:gaussian noise}a) and the TV-$L_1$ (\FullFigRef{fig:gaussian noise}b) and A$^2$TV (\FullFigRef{fig:gaussian noise}c) of \FullFigRef{fig:Recon_Params_Grid_TVL1_Gaussian-Noise}d and \FullFigRef{fig:Recon_Params_Grid_A2TV_Gaussian-Noise}e, respectively, which correspond to the regularization parameters that achieved the lowest MAD values. The middle panel of the figure (\FullFigRef{fig:gaussian noise}d-f) shows a magnification of 4 patches taken from the reconstructions, whereas the bottom panel (\FullFigRef{fig:gaussian noise}g) presents a 1D slice taken over the yellow line in \FullFigRef{fig:gaussian noise}b and \FullFigRef{fig:gaussian noise}c.  While both TV-$L_1$ and A$^2$TV significantly improved the reconstruction quality, in A$^2$TV more of the noise-induced texture between the blood vessels could be removed without damaging the structure of the blood vessels, thus leading to a lower MAD.}

\begin{figure}[hp!]
\centering
  \ifpdf
    \includegraphics[width=0.46\textwidth]{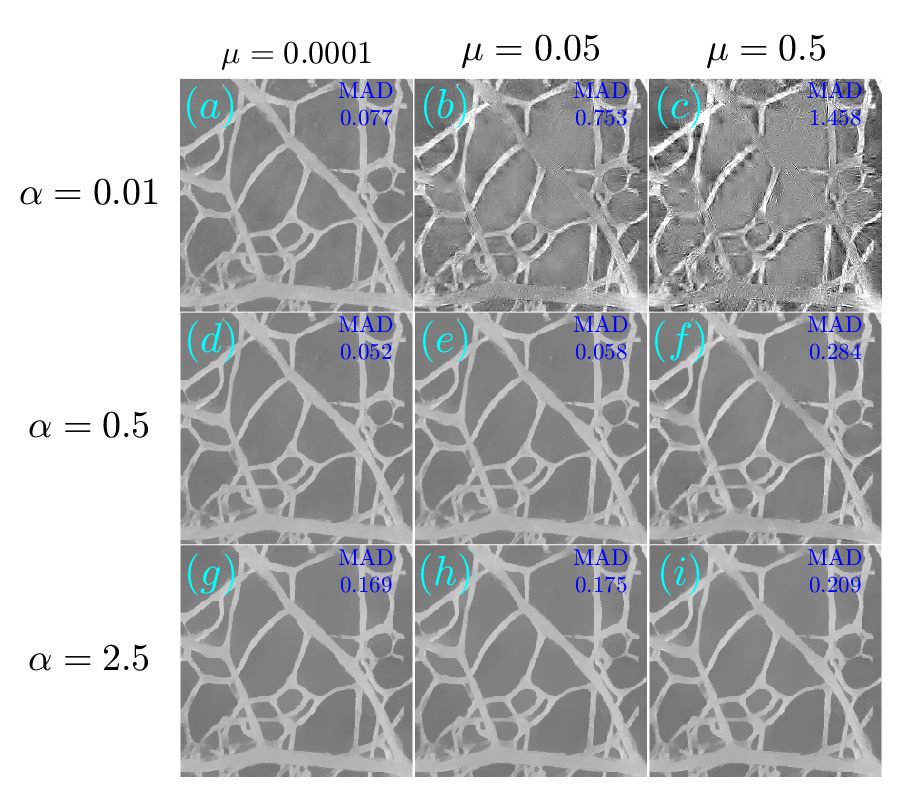}
  \else
    \includegraphics[width=0.46\textwidth]{Graphics/Optoacoustic_TVL1_256_lowres32.png}
  \fi
    \caption{\textcolor{black}{(a-i) The reconstruction of the image shown in \FullFigRef{fig:original_image}a for the case of under-sampled projection data using different parameters for the TV-$L_1$ case. The reconstructions were performed with 1000 iterations.}}
  \label{fig:Recon_Params_Grid_TVL1_LowRes32}
\end{figure}

\begin{figure}[hp!]
\centering
  \ifpdf
    \includegraphics[width=0.46\textwidth]{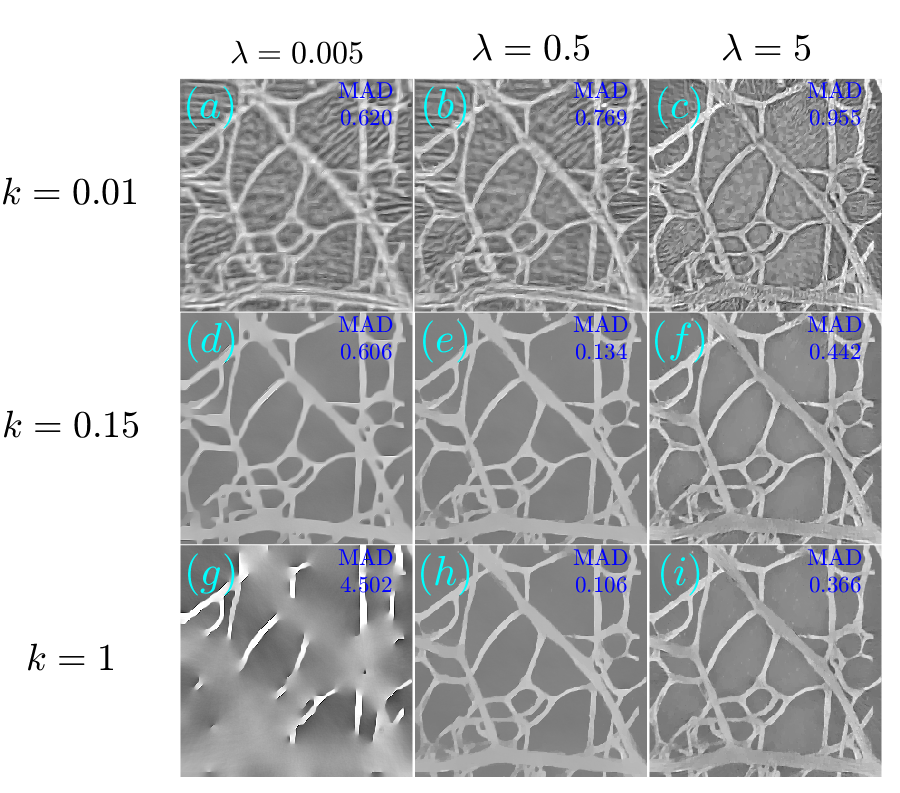}
  \else
    \includegraphics[width=0.46\textwidth]{Graphics/Optoacoustic_A2TV_256_lowres32.png}
  \fi
    \caption{\textcolor{black}{(a-i) The reconstruction of the image shown in \FullFigRef{fig:original_image}a for the case of under-sampled projection data using different parameters for the A$^2$TV case. The reconstructions were performed with 1500 iterations.}}  \label{fig:Recon_Params_Grid_A2TV_LowRes32}
\end{figure}

\begin{figure}[hp!]
\centering
  \ifpdf
    \includegraphics[width=0.46\textwidth]{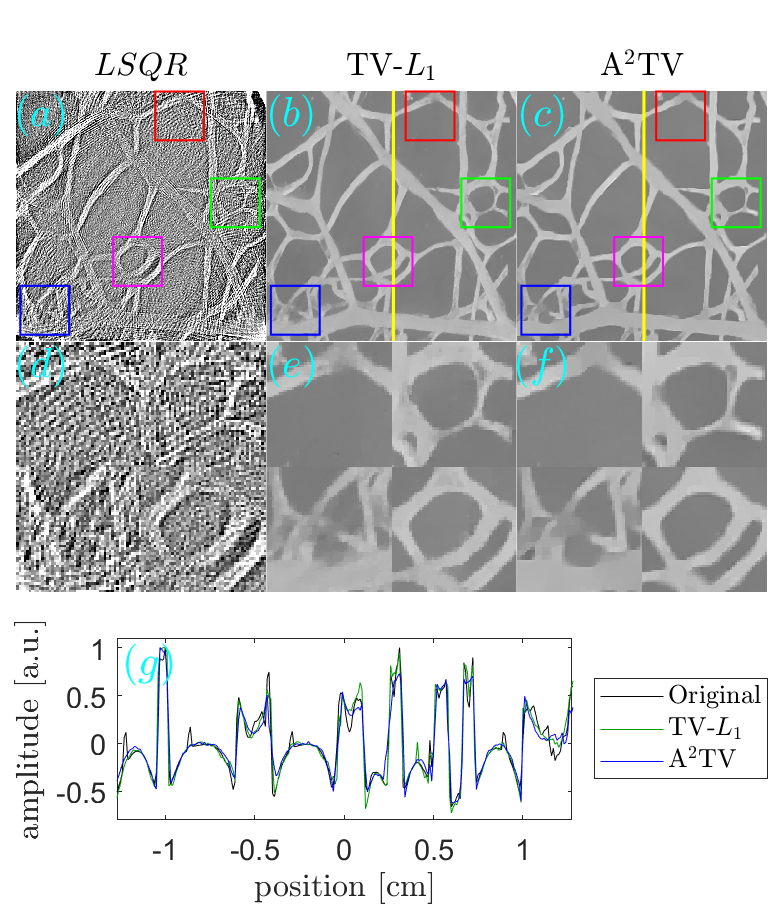}
  \else
    \includegraphics[width=0.46\textwidth]{Graphics/Optoacoustic_Hans_Comparison_lowres32_Noise_Example.png}
  \fi
  \caption{(a-c) The reconstruction of the image shown in \FullFigRef{fig:original_image}a for the case of under-sampled projection data using (a) LSQR, (b) TV-$L_1$, and (c) A$^2$TV. The bottom panel (d-f) is as in \FullFigRef{fig:gaussian noise}.  \textcolor{black}{The TV-$L_1$ and A$^2$TV reconstructions were produced using 1000 and 1500 iterations, respectively.} \textcolor{black}{(g) A 1D slice of the TV-$L_1$ and A$^2$TV reconstructions, corresponding to the vertical lines on the top panel, in comparison to a 1D slice taken from the original image.}}
  \label{fig:sparse projection}
\end{figure}

\textcolor{black}{The reconstructions for the case of 32 projections are presented in a similar manner to the case of noisy data. \FullFigRef{fig:Recon_Params_Grid_TVL1_LowRes32}, obtained with 1000 iterations, and \FullFigRef{fig:Recon_Params_Grid_A2TV_LowRes32}, obtained with 1500 iterations, respectively show the TV-$L_1$ and A$^2$TV reconstructions for a scan regularization parameters, whereas \FullFigRef{fig:sparse projection} shows a comparison between the LSQR and TV-$L_1$ and A$^2$TV reconstructions that achieved the lowest MAD (\FullFigRef{fig:Recon_Params_Grid_TVL1_LowRes32}d and \FullFigRef{fig:Recon_Params_Grid_A2TV_LowRes32}e, respectively). \textcolor{black}{All the A$^2$TV reconstructions were obtained with $\sigma=1.5$ pixels and $\rho=1$ pixel}. \FullFigRef{fig:sparse projection} shows that both TV-$L_1$ and A$^2$TV eliminated the streak artifacts that appeared with in the LSQR reconstruction, where the lowest MAD was achieved by the TV-$L_1$ reconstruction. Since both regularization methods eliminated the streak artifacts, the lower MAD achieved by TV-$L_1$ is a result of its ability to better preserve texture within the blood vessels, whereas in the A$^2$TV much of the blood-vessel texture was lost. Indeed, when examining the 1D slice in \FullFigRef{fig:sparse projection}g, it is easy to see that the TV-$L_1$  reconstruction captures the variations within the blood vessels better, whereas in the A$^2$TV reconstruction, these variations are smoothed.}

\textcolor{black}{In both the cases studied, A$^2$TV exhibited a higher capability than TV-$L_1$ to perform regularization without harming non-convex structures. Even in the case in which TV-$L_1$ achieved a lower MAD, the higher ability of A$^2$TV to preserve the fine details of the blood vessel morphology can be observed when comparing \FullFigRef{fig:sparse projection}e to \FullFigRef{fig:sparse projection}f. Additionally, in all the examples,  one can observe that when TV-$L_1$ was performed with a high level of TV regularization (bottom row in \FullFigRef{fig:Recon_Params_Grid_TVL1_Gaussian-Noise} and \FullFigRef{fig:Recon_Params_Grid_TVL1_LowRes32}) significant smearing of the blood vessels was observed. In contrast, in the A$^2$TV reconstructions, the smearing owing to over-regularization (low values of $\lambda$) could be diminished by increasing the anisotropy in the regularization, i.e. reducing the value of $k$. For the lowest values of $k$, higher levels of regularization (low values of $\lambda$) created undesirable anisotropic vessel-like texture in the reconstructed images.}

\section{Experimental Results}
\textcolor{black}{To further validate the suitability of A$^2$TV regularization for OAT image reconstruction, we tested its performance on experimental data. The optoacoustic setup comprised an optical parametric oscillator (OPO), which produced nanosecond optical pulses with an energy  30 mJ  at a repetition rate of 100 Hz and at a wavelength of $\lambda=680$ nm.  (SpitLight DPSS 100 OPO, InnoLas Laser GmbH, Krailling, Germany). The OPO pulses were delivered to the imaged object using a fiber bundle (CeramOptec GmbH, Bonn, Germany). Ultrasound detection was performed by a 256-element annular array (Imasonic SAS, Voray sur l'Ognon, France) with a radius of 4 cm, and an angular coverage of 270 degrees, comparable to the geometry shown in Fig. \ref{fig:simulation_illustration}. The ultrasound detectors were cylindrically focused to a plane, approximating a 2D imaging scenario.}

\begin{figure}[b]
\centering
  \ifpdf
    \includegraphics[width=0.3\textwidth]{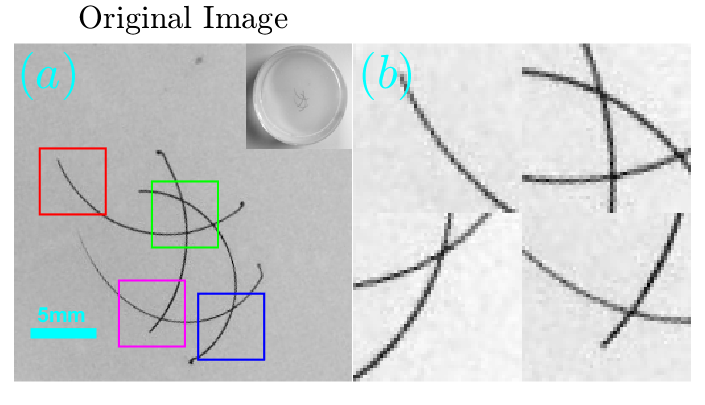}
  \else
    \includegraphics[width=0.3\textwidth]{Graphics/Optoacoustic_Experiment_OriginalImage.png}
  \fi
  \caption{(a) The originating image on which all the reconstructions were performed and (b) a binary mask that was generated from it. The bottom panel (c-d) shows magnifications of 4 marked regions in the respective images in the top panel.}
  \label{fig:original_Experiment_image}
\end{figure}

The imaged object was a transparent agar phantom which contained four intersecting hairs. \textcolor{black}{A photo of the phantom is shown in Fig.\ref{fig:original_Experiment_image}a, where Fig.\ref{fig:original_Experiment_image}b shows 4 magnified parts of the phantom.} The phantom preparation involved mixing 1.3\% (by weight) agar powder (Sigma-Aldrich, St. Louis, MO) in boiling water and pouring the solution in a cylindrical mold until solidification. To assure that all four hair \textcolor{black}{strands } \textcolor{black}{lie approximately in the same plane, we first prepared a clear cylindrical agar phantom, on which the hairs were placed; additional agar solution was then poured on the structure to seal the hairs. }

\begin{figure}[H]
\centering
  \ifpdf
    \includegraphics[width=0.46\textwidth]{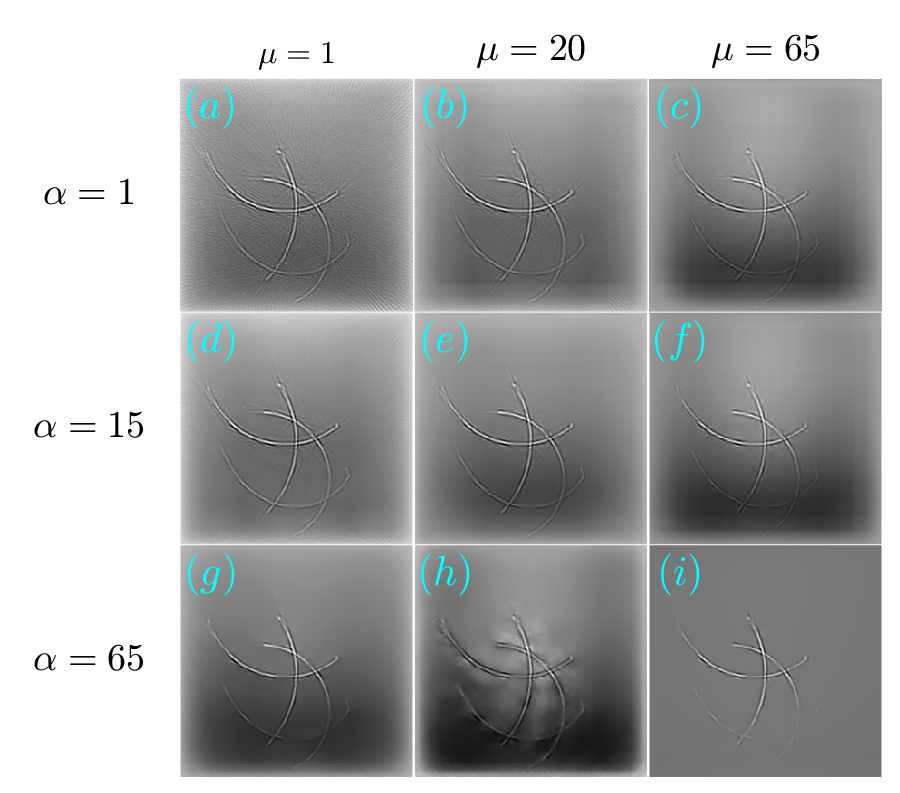}
  \else
    \includegraphics[width=0.46\textwidth]{Graphics/Optoacoustic_TVL1_256_experiment.png}
  \fi
    \caption{\textcolor{black}{(a-i) The reconstruction of the image shown in \FullFigRef{fig:original_Experiment_image}a for the case of experimental data using different parameters for the TV-$L_1$ case. The reconstructions were performed with 3000 iterations.}}
  \label{fig:Recon_Params_Grid_TVL1_Exp}
\end{figure}

\begin{figure}[H]
\centering
  \ifpdf
    \includegraphics[width=0.46\textwidth]{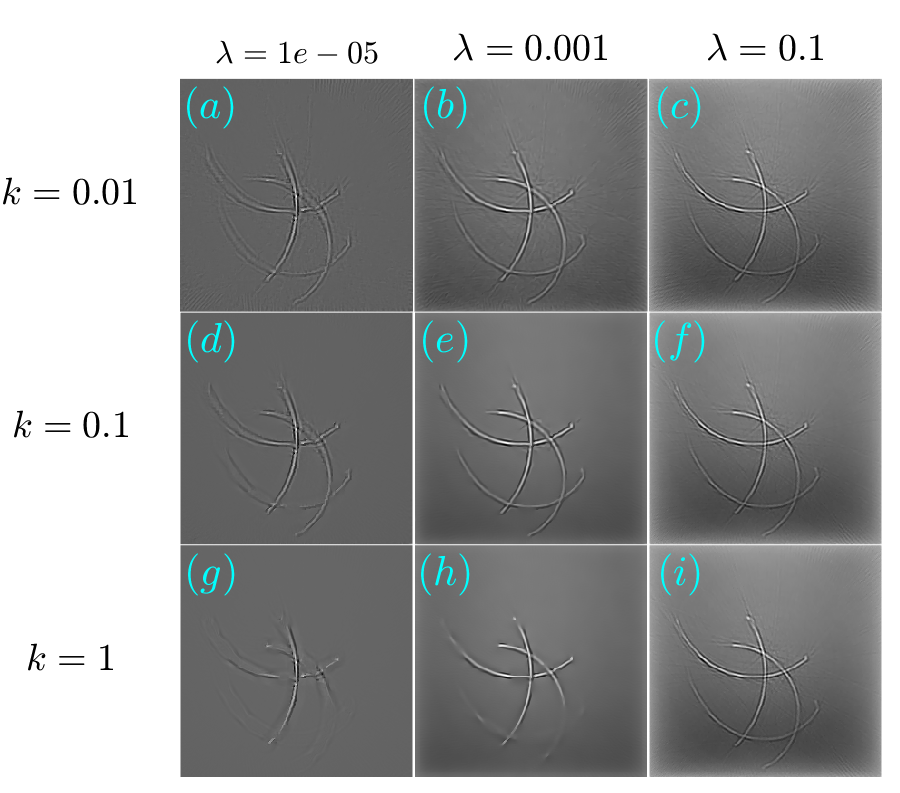}
  \else
    \includegraphics[width=0.46\textwidth]{Graphics/Optoacoustic_A2TV_256_experiment.png}
  \fi
    \caption{\textcolor{black}{(a-i) The reconstruction of the image shown in \FullFigRef{fig:original_Experiment_image}a for the case of experimental data using different parameters for the A$^2$TV case. The reconstructions were performed with 6000 iterations.}}
  \label{fig:Recon_Params_Grid_A2TV_Exp}
\end{figure}

\textcolor{black}{The optoacoustic image was reconstructed from the measured data using TV-$L_1$ regularization with 3000 iterations and A$^2$TV regularization with 6000 iterations, \textcolor{black}{ $\sigma=1.5$ pixels, and $\rho=1$ pixel.} Figs. \ref{fig:Recon_Params_Grid_TVL1_Exp} and \ref{fig:Recon_Params_Grid_A2TV_Exp} respectively show the images obtained via TV-$L_1$ and A$^2$TV reconstructions for a set of regularization parameters. As in the previous section, over-regularization in the TV-$L_1$ led to significant loss of structure, whereas for the A$^2$TV the ability to capture the image morphology under strong regularization (low $\lambda$) was improved when the anisotropy was increased via low values of $k$. }

\begin{figure}[H]
\centering
  \ifpdf
    \includegraphics[width=0.46\textwidth]{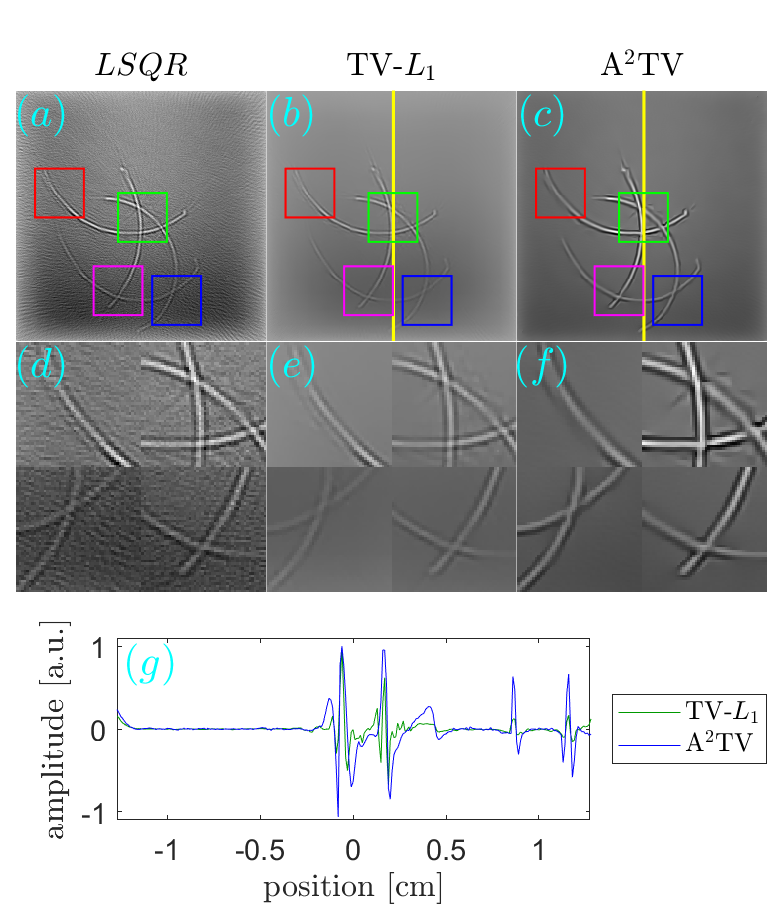}
  \else
    \includegraphics[width=0.46\textwidth]{Graphics/Optoacoustic_Hans_Comparison_experiment_Noise_Example.png}
  \fi
  \caption{(a-c) The reconstruction of the image shown in \FullFigRef{fig:original_Experiment_image}a for the case of experimental data using (a) LSQR, (b) TV-$L_1$, and (c) A$^2$TV. \textcolor{black}{The TV-$L_1$ and A$^2$TV reconstructions were produced using 3000 and 6000 iterations, respectively.} \textcolor{black}{(g) A 1D slice of the TV-$L_1$ and A$^2$TV reconstructions, corresponding to the vertical lines on the top panel.}}
  \label{fig:Experiment}
\end{figure}

\textcolor{black}{Fig.\ref{fig:Experiment} compares between the unregularized LSQR reconstruction (Fig. \ref{fig:Experiment}a) and the TV-$L_1$ (Fig. \ref{fig:Experiment}b) and A$^2$TV (Fig. \ref{fig:Experiment}c) reconstructions, respectively taken from Figs. \ref{fig:Recon_Params_Grid_TVL1_Exp}e and \ref{fig:Recon_Params_Grid_A2TV_Exp}e. The second row in the figure (Figs. \ref{fig:Experiment}d-\ref{fig:Experiment}f) shows a magnification of 4 patches from the three reconstructions of the top row (Figs. \ref{fig:Experiment}a-\ref{fig:Experiment}c), whereas the bottom panel (Fig.\ref{fig:Experiment}g) shows a 1D slice from the vertical yellow line in Figs. \ref{fig:Experiment}b and \ref{fig:Experiment}c. To allow an easy comparison, the 1D slices were normalized by their maximum values. We note that the negative values in the reconstruction are a common result of the limited detection bandwidth of the ultrasound detector \cite{rosenthal2013acoustic:5}. The figure clearly shows that the A$^2$TV reconstruction obtained the highest image quality, in particular for the weak hair structure at the image bottom. In particular, the 1D slice shows that the bottom hairs that appear around the position of 1 cm achieve a peak-to-peak signal over 4-times higher in the A$^2$TV reconstruction than in the TV-$L_1$ reconstruction.}




\section{Discussion}
In this work, a novel regularization framework was developed for OAT image reconstruction. The new
framework is based on an A$^2$TV cost function, which represents a generalization of the conventional TV functional, that is compatible with objects that possess complex, nonconvex boundaries. In
contrast to TV, the A$^2$TV cost function is an adaptive functional whose form depends on the
characteristics of the image. When using A$^2$TV, one first roughly defines the boundaries of the
objects in the image. Then, one uses these boundaries to determine the directions in which the
gradients are applied on the image. 

Similar to TV, A$^2$TV is most appropriate as a regularizer in optoacoustic image reconstruction when
the images are comprised of objects with well-defined boundaries. However, A$^2$TV is useful also
when these boundaries are incompatible with TV regularization due to their complexity. A common
category of optoacoustic images that fits the above description is blood-vessel images. Since blood is a major source of contrast in optoacoustic imaging, OAT systems often produce
images that are dominated by a complex structure of interwoven blood vessels. In particular, high-resolution images of the micro-vasculature are characterized by a complex network of arterioles, venues, and capillaries with extremely complex, nonconvex boundaries. 

\textcolor{black}{In our current implementation, A$^2$TV required setting 4 parameters: $\sigma$, $\rho$, $\lambda$, and $k$. The first two parameters, $\sigma$ and $\rho$, determined the image smoothing used in calculating the image gradients. While smoothing reduces the noise, thus limiting the effect of reconstruction errors on the detection of the principle axes, only moderate smoothing may be used without the risk of merging the boundaries of different objects in the image. Therefore, in all our examples, the smoothing was performed with Gaussian kernels whose standard deviations, $\sigma$ and $\rho$, were 3 pixels or less. While the choice of $\sigma$ and $\rho$ depended on the level of noise and artifacts in the regularization-free reconstruction, the parameters $\lambda$ and $k$ were determined by the amount of regularization desired in the reconstructed image, where $k$ determined the amount of anisotropy and $\lambda$ determined the strength of the regularizer. In our examples, performing over-regularization ($\lambda=0.0001$) led to image smearing in the isotropic case ($k=1$) and vessel-like artifacts in the case of high anisotropy ($k=0.01$). While such artifacts are undesirable, it is worth noting that their presence did not obscure the underlying image morphology, whereas over-regularization in TV-$L_1$ led to loss of image details.}

\textcolor{black}{We compared the performance of the A$^2$TV algorithm TV-$L_1$ algorithm for both numerically simulated data and experimental data. In the numerical simulations, an image of blood vessels was reconstructed for the cases additive noise and sparse sampling of the projection data. In the experimental reconstructions, the imaged object was four intersecting hair }\textcolor{black}{strands } \textcolor{black}{whose structure emulated the morphology of blood vessels.} \textcolor{black}{In both the numerical and experimental examples, A$^2$TV demonstrated a higher ability to preserve the blood-vessel morphology for high regularization parameters. Nonetheless, in the numerical example in which the reconstructions were performed with a low number of projections, TV-$L_1$ regularization achieved a lower MAD owing to the loss of texture in the A$^2$TV reconstruction. In the experimental example, A$^2$TV led to a considerable improvement in the contrast in the weak structures of the image in comparison to the TV-$L_1$ reconstruction.}

\textcolor{black}{The reconstruction performance demonstrated in this work suggests that A$^2$TV may be a useful tool for improving the ability of optoacoustic systems to perform vasculature imaging, which is a major application in the field. Since the texture within the blood vessels is affected by the random
distribution of the red-blood cells within the blood vessels, its elimination by A$^2$TV may be considered as an acceptable price for better visualization of the blood-vessel morphology.  In deep-tissue OAT systems, sub-millimeter vasculature imaging has been suggested as a potential diagnostic tool and has been demonstrated in the human extremities \cite{wray2019photoacoustic,matsumoto2018label} and breast \cite{toi2017visualization}. We note that while some OAT systems can also produce images of the tissue bulk, characterized by low frequencies and representative of the density of the microvasculature and fluence map, such systems require transducers capable of detecting ultrasound frequencies considerably below 1 MHz \cite{rosenthal2013acoustic:5}. In high-resolution OAT systems, operating at frequencies above 1 MHz and capable of reaching resolutions better than 100 $\mu$m \cite{dean2017optoacoustic,gateau2013ultra}, only signals from blood vessels may be detected. Finally, when performing optoacoustic imaging at resolutions better than 10 $\mu$m, e.g. using raster-scan optoacoustic mesoscopy (RSOM) \cite{omar2015pushing}, the image is dominated by the microvasculare and generally lacks any bulk component associated with blood.}


\textcolor{black}{We note that the formalism of A$^2$TV, in which the structure tensor matrix analysis is performed via eigenvalue decomposition \FullEqRef{Eq:StructureTensor_generalcase} enables its adaptation to higher image dimensions. TV regularization has been recently performed in 4D optoacoustic reconstruction that included 3 spatial dimensions and time \cite{ozbek2018optoacoustic}. Since the variation of the pixels in time is generally different than the one space, it may be expected that A$^2$TV can further improve image fidelity in such cases.}

\section{Acknowledgement}
We are thankful to Anna M. Randi for providing information on the image used in Section 4.

\appendix
\section{Solving the generalized ROF model}
\textcolor{black}{
A$^2$TV based reconstruction amounts to solving the following generalized ROF model,
\begin{equation}
\mathbf{u}^*=\arg\min_{\mathbf{u} \in X}\mathcal{E}(\mathbf{u})=\\ \arg\min_{\mathbf{u} \in
X}\mathcal{J}_{A^2TV}(\mathbf{u})+\frac{\lambda}{2}\|\mathbf{M}\mathbf{u}-\mathbf{p}\|_2^2.
\end{equation}
Where $\mathbf{M}$ is the model matrix, $\mathbf{p}$ is the acoustic pressure wave and $\mathbf{u}^*$ is the reconstructed image. 
The algorithm used for inversion is Chambolle-Pock, which solves the following primal form \cite{chambolle2011first:32}:}
\begin{equation}
    \arg\min_{\mathbf{u}\in X}F(K\mathbf{u})+G(\mathbf{u}).
\end{equation}

Where, in our case,
\begin{equation}
    \begin{array}{cc}F(\mathbf{y})=\|\mathbf{y}\|_1,G(\mathbf{u})=\frac{\lambda}{2}\|\mathbf{M}\mathbf{u}-\mathbf{p}\|_2^2\\
    K=\nabla_A=\textrm{grad}_A,K^*=\nabla_A^T=\textrm{div}_A \end{array}.
\end{equation}

Let X and Y be two finite-dimensional real vector spaces and $K$, a linear operator $K: X \rightarrow Y$ and its hermitian adjoint $K^*: Y \rightarrow X$. The appropriate saddle point problem is,
\begin{equation}
    \arg\min_{\mathbf{u}\in X}\arg\max_{\mathbf{z}\in
    Y}<K\mathbf{u},\mathbf{z}>+G(\mathbf{u})-F^*(\mathbf{z}).
\end{equation}
Where $\mathbf{z}\in Y$ is the dual variable. In our case,
\begin{equation}
    \arg\min_{\mathbf{u}\in X}\arg\max_{\mathbf{z}\in
    Y}<\nabla_A\mathbf{u},\mathbf{z}>+\frac{\lambda}{2}\|\mathbf{M}\mathbf{u}-\mathbf{p}\|_2^2-\delta_P(\mathbf{z}),
\end{equation}
where
\begin{equation}
    F^*(\mathbf{z})=\delta_P(\mathbf{z}),  
\end{equation}
and
\begin{equation}
    \delta_P(\mathbf{z})=
    \begin{cases} 
        0, & \mathbf{z}\in P \\
        \infty, & \mathbf{z}\notin P
    \end{cases}
    \quad , \quad  P=\{\mathbf{z}\in Y \ | \ \|\mathbf{z}\|_{\infty}\leq 1\}. 
\end{equation}
But, because of the appearance of the model matrix, we shall adjust the saddle point problem to
\begin{equation}
    \begin{split}
    \displaystyle
    &\arg\min_{\mathbf{u} \in X} \arg\smashoperator{\max_{\mathbf{z} \in Y, \mathbf{q} \in
    Q}}<\nabla_A\mathbf{u},\mathbf{z}>+<\mathbf{M}\mathbf{u}-\mathbf{p},\mathbf{q}>
    -\frac{1}{2\lambda}\|\mathbf{q}\|_2^2-\delta_P(\mathbf{z})\\
    &
    = \arg\min_{\mathbf{u} \in X}\arg\smashoperator{\max_{\mathbf{z} \in Y, \mathbf{q} \in
    Q}}
    <\nabla_A\mathbf{u},\mathbf{z}>+<\mathbf{M}\mathbf{u},\mathbf{q}>-<\mathbf{p},\mathbf{q}>\\ 
    &  \qquad   \qquad \qquad \qquad-\frac{1}{2\lambda}\|\mathbf{q}\|_2^2-\delta_P(\mathbf{z}) \\ 
    &=\arg\min_{\mathbf{u} \in X} \arg\smashoperator{\max_{\mathbf{z} \in Y, \mathbf{q} \in
Q}}< \begin{bmatrix} \mathbf{M} \\ \nabla_A \end{bmatrix} \mathbf{u},\begin{bmatrix} \mathbf{q} \\
\mathbf{z} \end{bmatrix} >-<\mathbf{p},\mathbf{q}>
-\frac{1}{2\lambda}\|\mathbf{q}\|_2^2-\delta_P(\mathbf{z}),
\end{split}
\end{equation}
using
\begin{equation}
\frac{\lambda}{2}\|\mathbf{M}\mathbf{u}-\mathbf{p}\|^2 =  \arg\max_{\mathbf{q} \in Q}
<\mathbf{M}\mathbf{u} - \mathbf{p},\mathbf{q}> - \frac{1}{2\lambda}\|\mathbf{q}\|^2.
\end{equation}
Here, the original saddle point problem is transformed into,
\begin{equation}
    \arg\min_{\mathbf{u} \in X}\arg\smashoperator{\max_{\mathbf{z} \in Y, \mathbf{q} \in
    Q}}<K\mathbf{u},\begin{bmatrix} \mathbf{q} \\ \mathbf{z} \end{bmatrix} 
    >+G(\mathbf{u})-F^*\left(\begin{bmatrix} \mathbf{q} \\ \mathbf{z} \end{bmatrix} \right).
\end{equation}
Where
\begin{equation}
\begin{split}F^*\left(  \begin{bmatrix} \mathbf{q} \\ \mathbf{z} \end{bmatrix}\right)
&= <\mathbf{p},\mathbf{q}>+\frac{1}{2\lambda}\|\mathbf{q}\|_2^2+\delta_P(\mathbf{z}), \\
G(\mathbf{u}) &= 0,\\  
K &= \begin{bmatrix}K_A\\K_{\nabla_A}\end{bmatrix} = \begin{bmatrix}\mathbf{M}\\
 \nabla_A\end{bmatrix} \\ 
 K^* & = \begin{bmatrix} K^*_A &  K^*_{ \nabla_A}\end{bmatrix} = \begin{bmatrix} \mathbf{M}^T
     & -\nabla_A^T \end{bmatrix}. 
     \end{split}
\end{equation}
Now, in order to use the algorithm, we shall use the proximal operator of $F^*$ and $G$,
\begin{equation}
\begin{split}
    \textrm{Prox}_{\sigma,F^*}\left( \begin{bmatrix}\tilde{\mathbf{q}} \\
        \tilde{\mathbf{z}}\end{bmatrix}  \right)&=\left(\frac{\tilde{\mathbf{q}}-\sigma
\mathbf{p}}{1+\frac{\sigma}{\lambda}},\left(\frac{\tilde{\mathbf{z}}}{\max
\left(1,|\tilde{\mathbf{z}}|\right)}\right)_{\textrm{pointwise}}\right),\\
    \textrm{Prox}_{\tau,G}\left(\tilde{\mathbf{u}}\right)&=\tilde{\mathbf{u}}
\end{split}
\end{equation}

\looseness=-1
\begin{algorithm}[t]
\begin{algorithmic}[1]
\Procedure{Chambolle-Pock}{$\textrm{Prox}_{\sigma_n,F^*}(\cdot),\textrm{Prox}_{\tau_n,G}(\cdot),K,K^*$}
	\State{Initialize $\gamma,\tau_0$}
    \State{$\sigma_0\leftarrow 1/(\tau_0L^2)$}
    \State{Initialize $u^0,q^0,z^0$}
    \State{$i\leftarrow 0$}
    \While{$i<N$}
    	\State{$[q^{n+1},z^{n+1}]\leftarrow \textrm{Prox}_{\sigma_n,F^*}(q^n+\sigma_nK_A(u^n),z^n+ \textcolor{white}{xxxxxx}  \sigma_nK_\nabla(u^n))$}
            	\State{$[u^{n+1},z^{n+1}]\leftarrow \textrm{Prox}_{\tau_n,G}(u^n-\tau_nK^*([q^{n+1},z^{n+1}]))$}

        \State{$\theta_{n+1}\leftarrow1/\sqrt{1+2\gamma\tau_n}$}
        \State{$\tau_{n+1}\leftarrow \theta_{n+1}\tau_n$}
        \State{$\sigma_{n+1}\leftarrow \sigma_n/\theta_{n+1}$}
        \State{Update tensor $A$ according to the new image $u^{n+1}$ \textcolor{white}{xxxxxx} and its dependents $K^*$ \& $K$}
    \EndWhile
	\State{return $u^N$}
\EndProcedure
\end{algorithmic}
\caption{ The Chambolle-Pock algorithm.}
\label{alg:1}
\end{algorithm}
The  algorithm is given in \FullFigRef{alg:1}, 
where in our case, we normalize the model matrix $\mathbf{M}$ by $\frac{1}{160}\sqrt{\|\mathbf{M}\|_\infty\|\mathbf{M}\|_1}$, where\\ $ \sqrt{\|\mathbf{M}\|_\infty\|\mathbf{M}\|_1}$ is an approximation to the model matrix $M$ Lipchitz constant. $L=L_{\mathbf{M}}+L_\nabla=160+8=168$, $\gamma =0.7\lambda$ and $\tau_0=0.5$

\section*{References}
\bibliographystyle{model1-num-names}
\bibliography{mybibfile}

\end{document}